\documentstyle[10pt,aaspp4]{article}
\begin{document}
\title{Persistent Counterparts to GRB}
\author{J. I. Katz\altaffilmark{1} and T. Piran}
\notetoeditor{Address correspondence to author at Department of Physics, 
Washington University, St. Louis, Mo. 63130 or katz@wuphys.wustl.edu.}
\affil{Racah Institute of Physics, Hebrew University, Jerusalem 91904, 
Israel}
\altaffiltext{1}{Permanent address: Department of Physics and McDonnell 
Center for the Space Sciences, Washington University, St. Louis, Mo. 63130}
\authoremail{katz@wuphys.wustl.edu}
\begin{abstract}
The recent discovery of persistent GRB counterparts at lower frequencies
permits several important conclusions to be drawn.  The spectrum of GRB970508
is not consistent with an external shock origin for both the prompt GRB and
the persistent emission, suggesting that at least the prompt radiation is
produced by internal shocks.  Comparisons
among three GRB with counterparts (or upper limits on them) establishes that
GRB are not all scaled versions of similar events.  The angular size 
inferred from the apparent observation of self-absorption in the radio 
spectrum of GRB970508 a week later implies that its expansion had slowed to
semi-relativistic speeds.  This permits a remarkably low upper bound to be 
placed on its residual energy, suggesting either that radiation has been 
$> 99.7\%$ efficient or that the initial outflow was strongly collimated.
Observations of self-absorbed radio emission from future GRB may permit
direct measurement of their expansion and determination of their parameters
and energetics.  We estimate initial Lorentz factors for GRB970228 and
GRB970508 $\gamma_0 \sim 100$, and present a solution for the evolution of a
blast wave with instantaneous cooling.  The visible flux is predicted to 
rise initially as a power of $t$ slightly less than unity, in agreement with
observations of GRB970508.
\end{abstract}

\section{Introduction}

The recent discovery by BeppoSAX and coordinated visible, infrared and
radio observations of persistent emission from gamma-ray bursts (GRB) has
answered several important questions.  Accurate coordinates of visible
counterparts have led to conclusive evidence that GRB are at cosmological
distances, definitively resolving the oldest and most fundamental question
in GRB astronomy.  Yet other questions, such as the nature of the GRB
emission mechanism, remain unanswered, and new questions have arisen.  Most
important, it is unclear how the persistent emission is produced and what
is its relation to the prompt soft gamma-ray radiation which has defined
GRB since their discovery in 1973.

In this paper we examine two different classes of models for the persistent
emission.  In one class it reflects continuing gamma-ray activity, not
directly observed because of the comparatively low sensitivity of GRB 
detectors to emission of low intensity but long duration.  In the other
class the persistent emission is the long-predicted afterglow produced
when a relativistically moving shell sweeps up ambient matter and its large
initial Lorentz factor decays as its energy is shared and radiated.  We
also examine two different classes of models for the prompt GRB emission.
In one it is produced by ``internal shocks''---interaction between
different shells each moving at different but relativistic speeds.  In the
other class it is the result of an ``external shock'' produced when a 
relativistic shell collides with the ambient matter.  We will argue that
the presently available observations (chiefly of GRB970111, GRB970228 and
GRB970508) present strong evidence that the prompt GRB emission is that
of internal shocks.  The explanation of the persistent emission is less
clear, with some evidence for continuing gamma-ray activity and some for
afterglows; it is possible that both processes contribute with
differing importance in different GRB.  We also present evidence that in
at least one burst (GRB970508) either the external shock is an extremely 
efficient radiator of the kinetic energy or the initial mass outflow and
gamma-ray emission are strongly collimated.

In \S2 we summarize the two classes of models for persistent emission.  
If the prompt GRB were produced by an external shock then in a burst whose 
persistent emission is afterglow it should be possible to extrapolate the 
properties of the afterglow back to the prompt GRB.  In \S3 we present 
evidence that this cannot be done for GRB970508, implying that its prompt 
emission was the result of internal shocks.  This conclusion is particularly
interesting because the prompt emission of GRB970508 had a simple 
single-peaked time profile, consistent with an external shock; external 
shocks are not satisfactory sources of the gamma-ray emission even where 
permitted by the time-history.  We compare three bursts in \S4 and conclude 
that GRB are not all scaled versions of similar events.  In \S5 we discuss 
evidence for self-absorption at radio frequencies in GRB970508, and estimate
the size of the emitting region.  From this we can infer either that 
radiation has been very efficient or that the initial expulsion of 
relativistic matter was strongly collimated.  Collimation differing from 
burst to burst is one possible explanation of the comparisons presented in 
\S4.  In \S6 we interpret the pulse shape and spectral history of GRB970228 
and GRB970508 as the results of internal shocks followed by external shocks,
and estimate their parameters.  \S7 presents simple estimates and scaling 
laws for the properties of afterglows produced when radiation is efficient, 
as suggested here.  \S8 contains a summary discussion and considers the 
prospects for determining the parameters of GRB and their afterglows by 
using measurements of self-absorbed synchrotron radiation to measure their 
expansion.  In an appendix we rederive and explain the result for the 
self-absorbed intensity of a GRB given (with little explanation) by 
\cite{K94a}.

\section{What Makes Persistent Emission?}

Gamma-ray bursts are generally agreed to be associated with the expulsion
of relativistic debris by a condensed object.  The fundamental argument
for this is the presence of gamma-rays with energies greater than $m_e c^2$.
If gamma-ray bursts are distant and luminous (the evidence for cosmological
distances and great luminosities is now compelling) such gamma-rays can
avoid destruction by $\gamma$-$\gamma$ pair production only if they are
narrowly collimated in a radially outward direction and their radius of
emission is large enough.  A sufficient, and probably necessary, condition 
for this to occur is that they be emitted by matter moving radially outward 
at relativistic speed with a Lorentz factor $\gg 1$.  

\subsection{Afterglow Models}

The outward moving matter will slow as a result of its  interaction
(generally assumed to be collisionless, for the collisional mean free path
is extremely long) with any surrounding medium.  Shocked matter will 
radiate, and as the Lorentz factor of the debris decreases the radiation
will shift to progressively lower frequencies and the corresponding time
scales will lengthen.

Considerations of this kind led several authors 
(\cite{K94a,K94b,MR97,SP97a,V97,W97}) to predict that GRB would be followed 
by an afterglow at frequencies gradually declining from soft gamma-rays 
through X-rays to visible light and radio waves.  It is possible to make 
many different assumptions concerning the efficiency of radiation, 
reconversion of post-shock thermal energy to bulk kinetic energy, 
collimation of the initial debris and development of the magnetic field.
\cite{MR93} and \cite{MRP94}, who integrated GRB spectra through the stages
which are now called afterglow, and \cite{PR93}, who considered radio
emission at very late times, demonstrate the possible variety and complexity
of these assumptions.  As a result, many different predictions of the 
quantitative development of the radiation spectrum with time are possible.

All such models in their basic form, in which only one set of assumptions
is made, have the general property that characteristic time scales and peak
intensities are power law functions of frequency.  The reason for this is
that if there is no characteristic Lorentz factor defined between the
Lorentz factor $\gamma_0$ of motion initially produced by the source and 
unity, then there is no opportunity for a break in the power law dependence
of any function on any variable between the values defined at these limits;
such a break would define a new characteristic Lorentz factor, which by
assumption is not present.  Of course, it is possible to define more complex
models in which this assumption does not hold (for example, radiation may be
assumed efficient in some range of Lorentz factors, and inefficient outside
that range).  This possibility opens the door to a plethora of parameters
and models of unmanageable complexity.  Such models may fit the data, but it
is hard to decide if this is because they are physically correct or because
it is easy to fit multiparameter functions of no physical significance; they
will not be considered further.

Afterglow models generally predict that the intensity at a given frequency
should rise until a characteristic time, and then decay (\cite{K94a,K94b}).
This appears to have been observed for visible light from GRB970508 
(\cite{CT97,D97,Ga97b,J97,Sch97}), although the initial rise is defined by 
only a very few observations.  The characteristic time is a decreasing 
function of frequency, short for gamma-rays, longer for visible light and yet 
longer at infrared and radio wavelengths.  Before this single maximum the 
intensity should rise monotonically, and afterwards it should decay 
monotonically, without fluctuation, as a smooth (usually power law, when well 
away from the maximum) function of time.  This is so far consistent with 
nearly all the data (see, however, \cite{Ko97}) but they are so sparse that 
this conclusion can hardly be said to have been proven.

\subsection{Continuing Emission Models}

In an alternative class of models (\cite{KPS97}) the persistent radiation at
lower frequencies (X-rays, visible light, infrared and radio waves) is 
produced by the same processes  which make the initial GRB.  The gamma-ray 
emission continues at lower intensity, perhaps maintaining a constant ratio 
to the lower frequency intensity.  

There is little direct evidence for or against models of this class.
Emission at frequencies below X-rays has never been observed simultaneously
with a GRB for instrumental reasons; only high upper bounds can be set on
its intensity.  Analogously, weak continuing gamma-ray emission is nearly
unobservable by a GRB detector, which is designed to trigger on intense
short bursts.  Continuing emission models agree with the spectral slope of 
persistent emission in I, V and X-ray bands in GRB970228 (\cite{KPS97}).
 
Some indirect evidence suggests that gamma-ray activity may continue long
past the nominal duration (usually $< 100$ s) of GRB.  High energy gamma-rays
were observed 5000 s after GRB940217 (\cite{H94}).  The ``Gang of Four''
spatially coincident (to within experimental accuracy) bursts of October 
27--29, 1996 (\cite{Me96,Co97}) indicate repetitive activity from a single 
source, which can equivalently be described as a single burst lasting two 
days, with brief periods of intense emission amidst a longer period of much
weaker or undetectably faint emission.

Models of this class predict that continuing gamma-ray activity should
accompany the lower frequency radiation.  In analogy to the behavior of
the intensity during observed shorter GRB, the intensity should fluctuate
irregularly in all bands, although this may be hard to detect in faint
objects for which observations require extended integrations.

\section{Evidence for Internal Shocks in GRB970508}

We wish to test the applicability of the models discussed in \S2 to GRB.
For a variety of reasons (satellite pointing constraints, instrumental
sensitivity, Earth occultation, daylight, moonlight, {\it etc.}) observations
of persistent emission by GRB are sparse.  In order to test continuing
emission models it is necessary to have data obtained simultaneously at
several frequencies.  The spectrum of GRB970508 does not support these 
models.

GRB970508 is therefore an excellent candidate for afterglow models.  All
the soft X-ray data and nearly all the visible data were obtained after the
peak intensities in those bands, and therefore reflect the behavior of a 
high energy tail to the particle distribution function (\cite{WRM97}) and 
not the evolution of its characteristic energy or gross energetics.  
Fortunately, if there are sufficient data to define the peak flux 
$F_{max\nu}$ at a frequency $\nu$ this quantity can be determined without 
the need for simultaneous measurements across the spectrum, or for frequent
sampling on both sides of the maximum.  Therefore, it may be easier to
compare $F_{max\nu}$ with theories than the instantaneous $F_\nu$ or than
$F_\nu(t)$ at a single $\nu$.

The data available for GRB970508 are shown in Figure 1.  The radio data are 
shown as a lower limit because (in the most recent available data;
\cite{F97b}) the flux is still rising.  The visible
(\cite{CT97,D97,Ga97b,J97,Sch97}), hard X-ray (\cite{C97b}) and gamma-ray 
(\cite{Kou97}) maxima were observed.  The soft X-ray flux (\cite{P97}) is a
lower limit because its maximum was missed.  Some specific afterglow models 
(\cite{K94a,K94b,MR97}) predict $F_{max\nu}$ to be independent of $\nu$; 
these are evidently excluded.
\placefigure{Fig1}

These points are not consistent with a single power law.  Simple afterglow
models (\S2.1) cannot describe the entire spectrum, regardless
of the specific assumptions made.  It is possible to resolve this problem
in many ways, for example by introducing two different regimes into the
afterglow model, with suitably chosen parameters.  We suggest, instead,
that the brief measured hard X-ray and gamma-ray emission are not produced 
by the same process as the persistent emission at lower frequencies.
\cite{SP97a} argued that the gamma-ray emission of multi-peaked GRB could not
be the result of an external shock, but rather was the result of internal
shocks.  We suggest that this conclusion is also applicable to GRB970508,
resolving the discrepancy with afterglow models, which may explain the
lower frequency emission.  Because GRB970508 was single-peaked in 
gamma-ray intensity (\cite{Kou97}) the arguments of \cite{SP97a} do
not apply to it directly.  Our argument for internal shocks in GRB970508
then suggests that in {\it all} GRB gamma-ray emission is produced by 
internal shocks, whatever the shape of their time profiles.

\section{GRB Differ Qualitatively}

It is natural to ask if all GRB are scaled versions of similar events.
External shock models predict that they are, because external shocks may
be scaled by powers of their basic parameters: distance, total energy,
initial Lorentz factor and ambient gas density.  Table 1 presents ratios of
$F_{max\nu}$ in four bands between GRB970508 and GRB970228.  These two 
bursts have quite different spectral properties.
\placetable{Tab1}
A similar conclusion may be reached by comparing GRB970508 to GRB970111
(Table 2). The different spectral properties are again evident.
\placetable{Tab2}

These tables summarize the obvious fact that GRB970508 was a comparatively
weak burst in gamma-rays.  It was brighter in V than GRB970228 and
brighter at 1.43 GHz than GRB970111, two bursts which were much stronger in
gamma-rays.  The obvious explanation is the same as that reached in \S3: 
the gamma-ray emission is the result of a different process than that which
produces the persistent emission at lower frequencies.

The weakness of GRB970508 in gamma rays poses other problems.  Its gamma-ray
fluence (\cite{Kou97}) and redshift (\cite{Me97}) imply an energy output
(if isotropic) of $\sim 3 \times 10^{51}$ erg.  This makes it difficult to 
argue that its comparatively low gamma-ray fluence results from unusually
inefficient gamma-ray emission, because if it had its observed V magnitude
and the same V/gamma-ray ratio as GRB970228 the observed fluence would
imply (assuming isotropy) the radiation of $\sim 10^{53}$ erg of gamma-rays.
This is excessive; it is unlikely (\cite{K97,KobPS97}) that internal shocks 
are even 50\% efficient!  

One possible explanation is that GRB970508 is, instead, an unusually
efficient source of visible and radio emission.  This would require that
the efficiency of lower-frequency afterglows varies greatly from burst to
burst, which seems implausible; {\it a priori}, one might expect more 
variation in the efficiency of a complex process like internal shocks,
which depend on the detailed temporal dependence of the Lorentz factor, than 
in that of a simple one like a blast wave, which only depends on a few
parameters.

A more promising alternative is that the gamma-ray emission of GRB970508 was
beamed towards us, so that the total power radiated was much less than that
implied by the assumption of isotropic radiation.  Beaming could solve all 
energetic problems, but the flux ratios require that other bursts show an 
even greater degree of gamma-ray beaming in order to explain their 
comparatively (to GRB970508) greater strength in gamma-rays in comparison 
to visible or radio fluxes.  Further evidence for beaming is discussed in 
the next section.

A third possible explanation is that discussed in \S2.2.  If the lower
frequency persistent emission (from soft X-rays down to radio waves) is
the product of continuing activity by the compact source of the gamma-ray
burst (\cite{KPS97}), then it would not be surprising if the strength of this 
activity differed from burst to burst, just as the time dependence of the 
gamma-ray emission differs.

\section{Self-Absorption, Radiative Efficiency and Beaming in GRB970508}

The observed (\cite{F97b}) radio spectrum of the persistent emission of
GRB970508 on May 15, 1997 was $\propto \nu^{1.1}$.  This is an
extraordinarily steep slope, and is naturally interpreted as the transition
between optically thick and thin conditions in the frequency range observed.
The expected self-absorbed spectrum (\cite{K94a}) is $\propto \nu^2$ 
(differing from the familiar $\nu^{5/2}$ of self-absorbed synchrotron 
radiation because in relativistic shocks the nearly all the low frequency 
radiation is emitted by electrons with a single, and much higher, 
characteristic synchrotron frequency).  The expected spectrum at frequencies
between the self-absorption frequency and the characteristic frequency is
$\propto \nu^{1/3}$.  

A theory for self-absorption in GRB was presented by \cite{K94a} (pp.
255--256) and is given in more detail in the Appendix.  The result is that 
below the self-absorption frequency
$$F_\nu = 2 \pi \nu^2 m_p \zeta (1 + z) {R^2 \over D^2}, \eqno(1)$$
where $R$ is the radius of the radiating shell, $z$ its redshift and $D = 
(2c/H_0)[1 - (1 + z)^{-1/2}]$ its proper distance.  Here we include the 
factor $\zeta \equiv k_B T_e / (\gamma m_p c^2)$ describing the degree of 
electron equipartition in the plasma shock-heated to an internal energy per 
particle $\gamma m_p c^2$ and moving with Lorentz factor $\gamma$.  Using 
the measured (\cite{F97b}) flux density on May 15, 1997 at 1.43 GHz as the 
optically thick value (this data point is only significant at the $2\sigma$ 
level, but is consistent with the very significant extrapolation of higher 
frequency flux densities) and taking $H_0 = 60$ km/s\,Mpc and $z = 0.835$
(\cite{Me97}) as the actual redshift yields $R \approx 4.1 \times 10^{16} 
\zeta^{-1/2}\ {\rm cm} \approx 1.2 \times 10^{17}$ cm for $\zeta = 1/9$ 
(a likely upper limit, assuming equal energy in electrons, ions and magnetic
field).  The usual result for the size after an interval $t$ is $R \sim 2 
\gamma^2 c t$, which would yield $\gamma \approx 2$, for which the 
relativistic approximation is barely valid.  \cite{S97} pointed out that 
because of the gradual deceleration of the shell the correct expression is 
$R \sim 8 \gamma^2 c t$ (if $\gamma \propto r^{-3}$, as in the blast wave 
model of \cite{K94a} and in \S7, this becomes $R \sim 14 \gamma^2 c t$), 
implying that the expansion is nonrelativistic by the time of these 
observations, a week after the burst.

This may be hard to reconcile with the inferred minimum energy $\sim 10^{52}$
erg because at a density of $n =1$ cm$^{-3}$ only $1.5 \times 10^{28}$ g 
($1.4 \times 10^{49}$ erg of rest mass energy) is contained within a sphere 
of $R = 1.3 \times 10^{17}$ cm.  In order for the motion to have been only 
marginally relativistic the kinetic energy must have been reduced to $\sim 
10^{49} n_1$ erg from its initial value, where $n_1 \equiv n/(1\,{\rm 
cm}^{-3})$, requiring a radiative efficiency $\epsilon > 99.7\%$, and 
probably at least 99.9\%.  Such efficient radiation is not possible from 
internal shocks for kinematic reasons (\cite{KobPS97,K97}), but may be in an
external shock (\cite{K94a,SNP96,S97}).  The hypothesis of extremely 
efficient radiation may be tested if a complete energy budget is obtained 
for future GRB, including the early stages of the external shock and the 
difficult spectral range of soft X-rays and the far ultraviolet; the early 
stages of the afterglow must emit as much radiation as the GRB itself, and 
perhaps much more.

Alternatively, if the relativistic debris shell is initially only produced
within a narrow solid angle $\Omega$ its energy requirement is reduced.
Once its motion becomes nonrelativistic the collimation is largely lost,
as shock-heated matter expands in all directions.  Then the inferred $\sim
10^{49} n_1$ erg of kinetic energy may represent most of the actual burst
energy, not requiring efficient radiative loss.  The measured GRB fluence,
now taken into the solid angle $\Omega$, implies a radiated energy of 
$\approx 2.5 \times 10^{50} \Omega$ erg.  Then
$$\Omega \approx 0.04 {\epsilon \over 1 - \epsilon} n_1. \eqno(2)$$
In internal shock models plausible values of $\epsilon \sim 0.5$ imply a
collimation opening angle $\theta \sim 0.1 n_1^{1/2}$ radian.

This degree of collimation requires $\gamma_0 \gtrsim
1/\theta \approx 10 n_1^{-1/2}$, consistent with other estimates.  It also 
argues against $n_1 \ll 10^{-3}$, such as would be found in galactic halos 
or intergalactic space, for this would lead to excessively small $\theta$; 
even $n_1 = 10^{-3}$, as found in much of the volume of the Galactic disc 
would imply $\gamma \gtrsim 300$.

\section{Possible Parameters of GRB970228 and GRB970508}
\subsection{GRB970228}

The gamma-ray (40--600\,keV) light curve of GRB970228 displays a strong and
rapidly variable peak about 4 seconds long, followed by a much weaker and
smoother peak that begins about 40 seconds later (\cite{C97a}). The first
pulse is variable on a time scale of $\sim 0.1$ s (\cite{H97}). The 
X-ray (2--30keV) light curve of GRB970228 displays two comparable peaks, 
corresponding in time to the gamma-ray peaks.  The observed flux in the 
second X-ray peak is comparable to the flux obtained from backward 
extrapolation of the persistent X-ray emission light curve.

This suggests a simple interpretation: The first GRB pulse was produced by 
internal shocks while the second pulse signals the beginning of an afterglow
produced by an external shock when the debris sweeps up ambient gas.  
Behavior of this kind was suggested by \cite{SP97a} who pointed out that 
internal shocks could not convert the total kinetic energy to radiation and 
that a significant fraction of the kinetic energy should be emitted later in
other parts of the spectrum.  This interpretation allows us to estimate some
of the parameters of the event.  

The overall duration of the first pulse corresponds to the duration of 
intense activity of the ``inner engine'' that emitted the ejecta. The 
variability on a time scale of $\sim 0.1$ s corresponds to variability in 
the ``inner engine'' and sets an (unsurprising) direct upper bound on its 
size of $\sim 3 \times 10^9$ cm.

The second pulse corresponds to the the onset of significant radiation by 
the external shock.  This takes place when the ejecta (with energy $E$,
Lorentz factor $\gamma_0$ and proper mass $M = E/\gamma_0 c^2$) sweeps up an 
ambient mass $M/\gamma_0$ at $r = R_\gamma \equiv [3 E/(4 \pi n m_p c^2 
\gamma_0^2)]^{1/3} \approx 1.2 \times 10^{18}\gamma_0^{-2/3} E_{52}^{1/3} 
n_1^{-1/3}\,{\rm cm}$, where $E_{52} \equiv E/(10^{52}\,{\rm erg})$.  The 
corresponding time of arrival of photons from the onset of the external 
shock is $t_e \sim R_\gamma/(2c \gamma_0^2) \approx 2 \times 10^7 
\gamma_0^{-8/3} E_{52}^{1/3} n_1^{-1/3}\,$s.  Comparison with the time delay
of 40 s yields $\gamma_0 \approx 140 (E_{52}/n_1)^{1/8}$. This is a 
reasonable value, but the dependence on $E$ and $n$ is so weak that
no significant constraints can be placed on them.  \cite{FEH93,WL95,P96} 
give lower limits to $\gamma > 100$ but \cite{SP97b,K97} give also an 
upper limit to $\gamma$ of the order of $10^3$ for the observation of 
internal shocks.

\subsection{GRB970508}

The observation of a visible maximum 1--2 days after GRB970508 
(\cite{CT97,D97,Ga97b,J97,Sch97}) suggests that at this time the characteristic
synchrotron frequency of the afterglow passed through the visible spectrum.
It should be possible to extrapolate this frequency backwards in time to $t
= t_e$, when $\gamma \approx \gamma_0$ and the afterglow begins.

In principle, this earlier part of the afterglow could have been observed
by continuously monitoring GRB detectors.  The fact that it was not seen
in GRB970508 (in contrast to the suggestion made above for 
GRB970228) suggests that its peak frequency was too low for observation by 
GRB detectors, although we cannot be sure (without quantitative estimates of
its radiative efficiency) how intense it would have been.  

Interpreting the nondetection by GRB detectors to mean that the peak 
characteristic afterglow photon frequency $\nu < \nu_0$, we find 
$\gamma_0 \lesssim 60 \zeta_e^{-1/2} \zeta_B^{-1/8} n_1^{-1/8} (h\nu_0 / 100
\,{\rm KeV})^{1/4}$.  Although this may be less than lower limits estimated 
by consideration of gamma-gamma pair production (\cite{FEH93,WL95,P96}), 
these limits are smaller for afterglow at larger radii than for prompt GRB 
emission, and are not directly applicable to a burst (such as GRB970508) for 
which no data exist on the high energy spectrum.  It is also plausible that
$\zeta_e$, $\zeta_B$ or $n_1$ are significantly less than unity, raising the
upper bound on $\gamma_0$.

At the time of the visible maximum we can estimate $\gamma \approx 4$ 
(Equation 7).  Extrapolation (using $\gamma \propto t^{-3/7}$; \S7) implies 
the blast wave will become only semirelativistic ($\gamma < 2$) after about 
a week, in agreement with the inference made in \S5.  

The implied (using scaling laws from \S7) X-ray ($h\nu \sim 10\,$KeV) maximum 
occurred $\sim 10^3\,$s after the burst, before the BeppoSAX follow-up 
observations (\cite{P97}) but well after the gamma-ray burst itself 
(\cite{Kou97}).  For this reason the X-ray emission measured simultaneously 
(\cite{C97b}) with the GRB was probably the low frequency extrapolation of 
the gamma-rays, produced by internal shocks, and not the beginning of the 
afterglow.  The simultaneous X-ray emission also had $F_{max}$ much greater
than the visible $F_{max}$, inconsistent with the theory for both weakly
(\cite{K94a,K94b}) and strongly (\S7) radiating blast waves, but consistent
with the downward ($\propto \nu^{1/3}$; \cite{K94a,K94b,Coh97}) extrapolation
of the gamma-ray spectrum.

\section{Blast Waves with Instantaneous Cooling}

Previous relativistic blast wave models 
(\cite{MR93,K94a,K94b,MRP94,MR97,V97,W97}) have generally assumed that only
a small fraction of the energy is radiated, as in the Taylor-Sedov 
nonrelativistic blast wave theory.  The arguments of \S5 suggest that this
may not be the case, at least in GRB970508, but that radiation may be very
efficient, as indicated by the estimated short electron radiation times
(\cite{K94a,SNP96}).   Alternatively, the angular spreading of an initial
narrowly collimated outflow as its Lorentz factor degrades will reduce the
energy per solid angle with increasing distance along the axis of the flow 
in a manner qualitatively resembling the effects of efficient radiation.

Here we develop a simple model for the evolution of a relativistic blast 
wave and its radiated spectrum, assuming that all the energy produced by the
shock is instantly radiated.  This is the limiting case opposite to that of 
negligible radiation, previously assumed.  Strong coupling between the 
protonic and the electronic thermal energies is required.

We describe the interaction of a relativistic blast wave with the ambient
matter by a series of inelastic collisions between the debris and previously
swept-up mass with proper mass $M + m$ and Lorentz factor $\gamma(r)$ and 
infinitesimal shells of proper mass $dm$.  The internal energy produced in 
each collision is instantaneously radiated.  Conservation of energy and 
momentum yield\footnote{Equations (71) and (72) of Katz (1994a) are wrong.}:
$${d\gamma \over \gamma^2 -1} = - {dm(r) \over M+m(r)}, \eqno(3)$$
which can be integrated:
$${\gamma + 1 \over \gamma - 1} = {\gamma_0 + 1 \over \gamma_0 - 1} \left(
1 + {m(r) \over M}\right)^2. \eqno(4)$$
Then
$$\gamma(r) = {(\gamma_0 + 1)[(r/L)^3 + 1]^2 + (\gamma_0 - 1) \over (\gamma_0
+ 1)[(r/L)^3 + 1]^2 - (\gamma_0 - 1)}, \eqno(5)$$
where $L \equiv
[3 M /(4 \pi m_p n)]^{1/3} = [3 E/(4 \pi m_p c^2 n \gamma_0)]^{1/3}$.  $L$ 
corresponds to the radius within which the ambient mass equals the proper 
mass of the debris. $L$ is smaller than the Taylor-Sedov radius $R_{TS}$ 
within which the ambient rest-mass energy equals $E$ and it is larger than 
$R_\gamma$; the ratios $R_{TS}:L:R_\gamma = \gamma_0^{2/3}:\gamma_0^{1/3}:1$.

There are three limits for this solution:
$$\gamma(r) \approx \cases{\gamma_0 & for $r \ll R_\gamma$ \cr 
(r/L)^{-3} & for $R_\gamma \ll r \ll L $ \cr 
1 + 2(r/L)^6 & for $L \ll r $ \cr} \eqno(6)$$ 
Comparison of this solution with the Lorentz factor when there is no
significant cooling reveals, naturally, that this Lorentz factor is smaller.
For our purposes the intermediate range is the most interesting and we will 
use it in the following discussion.  The $\gamma \propto r^{-3}$ dependence 
found here is also that found if it is assumed the blast energy is 
distributed uniformly, or in a way which scales with $r$, as $r$ increases 
(\cite{K94a}), and a similar assumption is made in the Taylor-Sedov
nonrelativistic blast wave theory.

This Lorentz factor is also the typical ``thermal'' Lorentz factor of
the relativistic shocked protons. The thermal Lorentz factor of the 
relativistic electrons is, by definition of the equipartition factor 
$\zeta_e$, $\gamma_e = \zeta_e (m_p/m_e) \gamma$.  The characteristic 
synchrotron radiation frequency from the forward shock (the backward shock
crosses the shell and is gone before the shell enters the $\gamma \ll 
\gamma_0$ regime) is (\cite{P94})
$$h\nu \approx 0.01\,{\rm eV}\,\zeta_e^2 \zeta_B^{1/2}\gamma^4 n_1^{1/2} 
\approx 0.01\,{\rm eV}\,\zeta_e^2 \zeta_B^{1/2} 
\left({r \over L}\right)^{-12} n_1^{1/2}, \eqno(7)$$
where the magnetic equipartition factor $\zeta_B$ is the ratio between the 
magnetic energy density and the ``thermal'' energy density of the shocked
protons.  For typical parameters the characteristic frequency is in the 
visible range when $\gamma \approx 4$, and the blast wave ceases to be 
relativistic when this frequency is in the infrared.

The energy emitted at radius $r$ and Lorentz factor $\gamma(r)$ will be 
observed around a time:
$$t(r) = {r/c \over 14\gamma^2 (r)} = (L/14c) (r/L)^{7}. \eqno(8)$$
Here we have replaced the usual factor of 2 in the denominator by 14 to 
allow for deceleration (\cite{S97}; see \S5).

Combining these results and setting the equipartition factors to unity we 
find:
$$h \nu \approx 0.00011\,{\rm eV}\,(ct/L)^{-12/7} n_1^{1/2} \approx
0.14\,{\rm eV}\left({t \over 1\,{\rm day}}\right)^{-12/7} E_{52}^{4/7} 
\left({\gamma_0 \over 300}\right)^{-4/7} n_1^{-1/14}.  \eqno(9)$$
This gives a scaling law $t \propto \nu^{-7/12}$ which may be used to 
compare or predict the times of peak flux at different frequencies.
Note also that $\gamma \propto t^{-3/7}$ and $r \propto t^{1/7}$.

We can now derive the scaling of $F_{max\nu}$ and $F_{\nu}(t)$, using 
Equations (6)--(8): The time integrated fluence per unit frequency $dE/d\nu 
= (dE/dr)(dr/d\nu) \propto r^9$.  This is emitted over a characteristic time
$\propto r^7$, yielding
$$F_{max\nu} \propto r^2 \propto \nu^{-1/6}. \eqno(10)$$
This should be valid throughout the regime in which the shell is moving
relativistically, roughly corresponding to $\nu$ from the infrared to 
X-rays.  This law can, in principle, be compared with data like those
shown in Figure 1, but in that Figure there is only one significant point
in the applicable frequency range.

At lower frequencies the instantaneous spectrum $F_{\nu} \propto \nu^{1/3}$
(\cite{K94b}), so that at a fixed frequency 
$$F_{\nu}(t) \propto F_{max\nu} \left({\nu \over \nu_{max}}\right)^{1/3}
\propto \nu^{1/3} t^{6/7}.  \eqno(11)$$
The exponent of $t$ is very close to the value of $4/5$ obtained in a model
in which radiation is inefficient $\cite{K94a,K94b}$.  Observations of the
persistent visible counterparts to GRB970228 (\cite{Gu97}) and GRB970508 
(\cite{D97}) confirm the predicted rise before the maximum, but are not
sufficient to determine the exponent quantitatively.  The data for
GRB970508 suggest an exponent of $0.7 \pm 0.1$, approximately consistent
with the predictions, but this value may underestimate the true exponent
because it is based on only two points, the second of which was obtained 
close to the maximum when the initial rise may have been leveling off.

\section{Discussion}

The persistent counterparts of GRB may offer as much complexity and
variety as the classical GRB themselves.  Future observations will provide
a catalogue of their behavior, and will show if this is true, or if they
may be described by a simple general model.  Such a general description,
with a few parameters to describe the possible collimation and degree of
energy loss, as well and $E$, $n$ and $\gamma_0$, would support afterglow
models, while complex ``no two of them alike'' behavior would support the
suggestion of continuing emission.  Afterglows are almost inevitable, and
were predicted by many authors, but there is also evidence for continuing
emission.  Perhaps each may be important in different bursts.

The apparent discovery of self-absorption in GRB970508 and the measurement
of its redshift offer the prospect of detailed reconstruction of the history
of a GRB blast wave.  It may be possible to measure directly the blast wave
radius as a function of time, and therefore to determine its slowing
history.  If radiation is efficient then by constructing an energy budget 
from the observed radiation it would be possible to determine $E$ directly.
By comparing the radiated power to the slowing history of the blast wave it
would then be possible to determine $n$.  It may also be possible to 
determine $\gamma_0$ (\S6,7).

If GRB are collimated, as suggested for GRB970508 in \S5 and has been
widely speculated on the grounds that their likely sources, coalescing
compact binaries, are far from spherically symmetric, the analysis of 
afterglows will be much more complicated.  They will have several more 
parameters, and will probably require numerical simulation.  However, even 
the first clear demonstration of collimation would be significant.

In Equation (8) most of the sensitivity of $t$ to $r$ comes from the
$\gamma^2$ denominator; to a fair approximation $t \propto \gamma^{-2}$.
Combining this with $\nu \propto \gamma^4$ (Equation 7) implies that $t$ 
will generally be proportional to approximately the -1/2 power of $\nu$.  We
expect this to hold for all fireball afterglow models, whatever their 
detailed physics.  For example, this paper found $t \propto \nu^{-7/12}$ and
the very different model of \cite{K94a,K94b} found $t \propto \nu^{-5/12}$.
If $\gamma \propto r^{-3/2}$, as assumed in many models, then $t \propto
\nu^{-2/3}$.  In simple synchrotron radiative cooling models without 
hydrodynamics $t \propto \nu^{-1/2}$.  For similar reasons the exponent in
Equation (11) is generally slightly less than unity.  This suggests that it 
may not be possible to distinguish among fireball afterglow models on the 
basis of the time dependence of the characteristic frequency or the 
pre-maximum rate of rise.  However, these are not the only means available;
for example, an energy budget will decide if radiation is energetically
important.

\acknowledgements

We thank Lifan Wang and Re'em Sari for discussions, Washington University 
for the grant of sabbatical leave, the Hebrew University for hospitality and
a Forchheimer Fellowship, and the U. S.-Israel BSF, NASA NAG 52682 and 
NAG5-1904 and NSF AST 94-16904 for support.

\appendix
\section{Self-Absorption}

A quantitative calculation of the self-absorbed flux from a rapidly moving
emitter is not trivial, because a retarded-time calculation must be performed
to allow for the fact that different points on the radiating surface are
observed at different radii and retarded times.  Deceleration and limb
darkening introduce further complications.  Here we only demonstrate that
for shock-heated emitters the Lorentz factor does not appear in the
expression for the observed flux density.

In the co-moving frame of the shocked radiating matter moving with Lorentz
factor $\gamma \gg 1$ the temperature is given by $T = \zeta f m_p c^2/k_B$,
where the Doppler shift factor $f = (1 + v \cos\theta/c)\gamma \approx 2 
\gamma$ (\cite{K94a} took $\zeta = 1$ and used a variable $\gamma_F$ which is
here equal to $f^2$).  In that frame the Rayleigh-Jeans brightness
$$B_\nu = {2 \nu^2 \over c^2} \zeta f m_p c^2. \eqno(A1)$$
The observer measures a blue-shifted temperature $\zeta f^2 m_p c^2/k_B$,
and hence observes a brightness
$$B_\nu^\prime = 2 \nu^2 \zeta f^2 m_p. \eqno(A2)$$
However, the observer sees significant radiation only from a solid angle
$\approx \pi r^2/(f^2 D^2)$ because of the relativistic beaming of radiation
into a cone of angle $1/f$.  The observed intensity
$$F_\nu \approx 2 \pi \nu^2 \zeta m_p {r^2 \over D^2}; \eqno(A3)$$
the powers of $f$ cancel.

An alternative derivation begins by noting that the quantum occupation
number $n_\nu$ is a Lorentz-invariant scalar.  The intensity
$$I_\nu = \nu^3 n_\nu.  \eqno(A4)$$
When Doppler shifting to the observer's frame $I_\nu \to I_\nu f^3$.
In the Rayleigh-Jeans limit the occupation number
$$n_\nu = {k_B T \over h \nu}. \eqno(A5)$$
The temperature $T \propto \gamma \propto f$ (not because of a Doppler 
shift, but from the shock jump conditions), so that $I_\nu \propto f^4$.
This result must be divided by $f^2$ to allow for the fact that the observed
radiation was emitted at a frequency lower by a factor $f$, and $I_\nu
\propto \nu^2$.  Finally, the observed effective radiating area is $\propto 
f^{-2}$, leading to the result that $F_\nu$ is independent of $\gamma$.

Each of these derivations uses the shock jump conditions.  The final result
cannot be obtained from Lorentz transformations of the radiation field 
alone.

\newpage
\figcaption[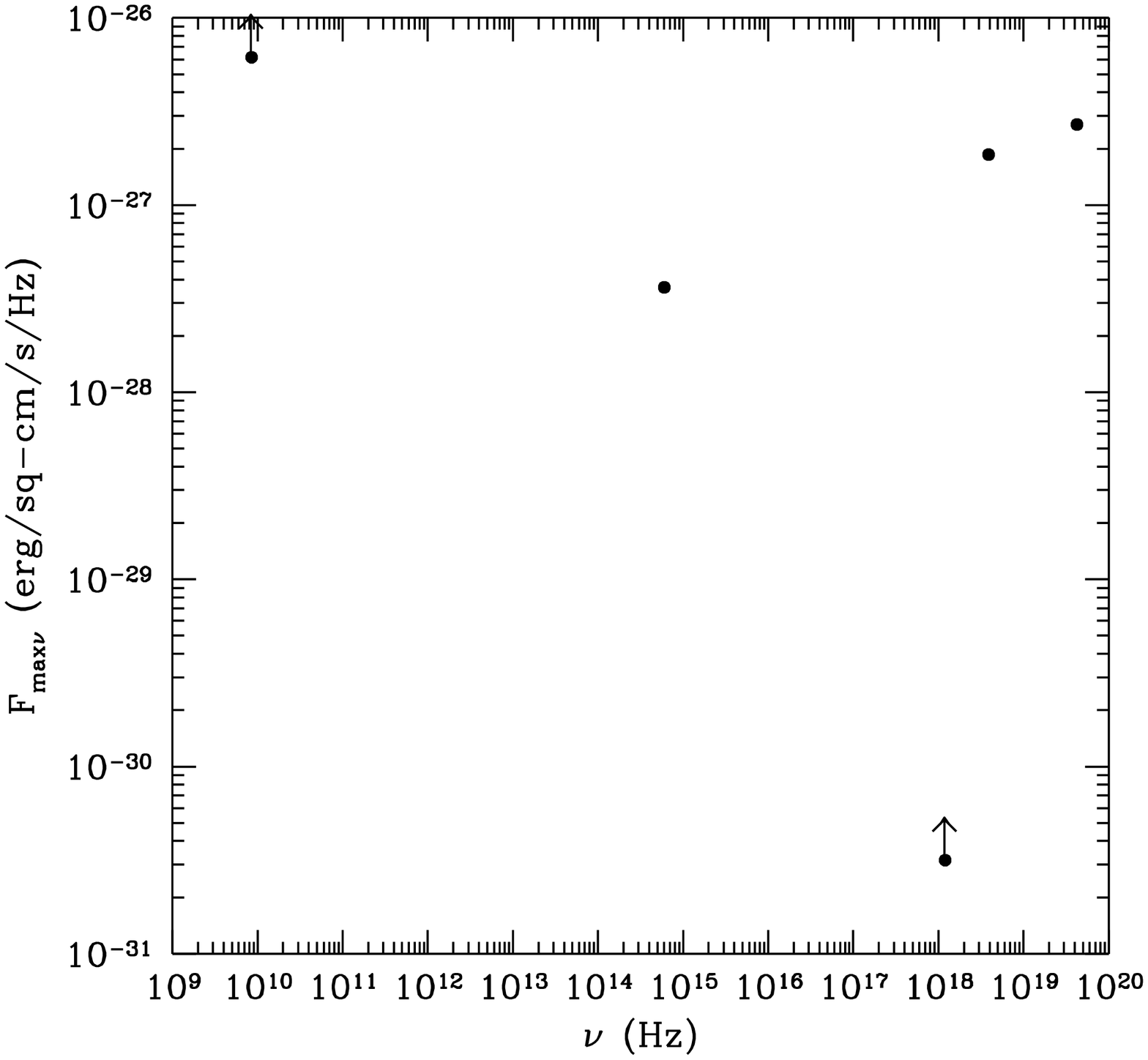]{Peak fluxes in four bands for GRB970508.  The radio
(8.46 GHz) and soft X-ray points are lower bounds, indicated by arrows, 
because the maxima in these bands were not observed.  The radio, visible and 
gamma-ray points are not consistent with a single power law.  Data are from 
Kouveliotou {\it et al.} (1997) (gamma-rays), Costa {\it et al.} 1997b (hard
X-rays), Piro {\it et al.} (1997) (soft X-rays), Galama {\it et al.} (1997b)
and Jaunsen {\it et al.} (1997) (V) and Frail {\it et al.} (1997b) (radio).
The X-ray and gamma-ray points depend significantly on assumptions about
the spectrum in those bands.
\label{Fig1}}
\newpage
\begin{table}
\centering
\begin{tabular}{cccc}
gamma-ray & 2--30 KeV & V & Radio \\
0.06 & 1.3 & $\sim 2$ & ? 
\end{tabular}
\caption{Ratios of peak fluxes, GRB970508 to GRB970228, in four bands.  Data
from Kouveliotou {\it et al.} (1997) (extrapolated to match BeppoSAX 40--600
KeV band) and Hurley {\it et al.} (1997) (gamma-rays), Costa {\it et al.} 
(1997ab) (2--30 KeV X-rays) and from Galama {\it et al.} (1997b), Jaunsen 
{\it et al.} (1997) and Groot {\it et al.} (1997) (V).  The entry under V 
assumes that the characteristic time scale and post-maximum decay rate of 
GRB970228 (whose maximum was not observed) is similar to that of GRB970508;
 the data directly imply a V band ratio of $< 3$. \label{Tab1}}
\end{table}
\begin{table}
\centering
\begin{tabular}{cccc}
gamma-ray & X-ray & V & 1.43 GHz \\
0.05 & ? & ? & $> 0.2$
\end{tabular}
\caption{Ratios of peak fluxes, GRB970508 to GRB970111, in four bands.  Data
from Kouveliotou {\it et al.} (1997) (extrapolated to match BeppoSAX 40--600
KeV band) and Frail {\it et al.} (1997a) (gamma-rays) and from Frail {\it et
al.} (1997ab) and Galama {\it et al.} (1997a) (radio).  A lower bound is 
given at 1.4 GHz because the radio flux of GRB970508 was still rising in 
the most recent available data. \label{Tab2}}
\end{table}
\newpage
\begin{figure}
\plotone{persistf.ps}
\end{figure}
\end{document}